\documentclass[twocolumn,english,aps,prl,twocolumn,superscriptaddress,nofootinbib]{revtex4}

\usepackage[T1]{fontenc}
\usepackage[latin9]{inputenc}
\usepackage{textcomp}
\usepackage{amssymb}
\usepackage{amsmath}
\usepackage{graphicx}
\usepackage{esint}
\usepackage{tikz}
\usepackage{subfigure}
\usepackage{hyperref}
\usepackage{bbold}
\usepackage{babel}

\usepackage{bm}
\usepackage{braket}
\makeatletter

\@ifundefined{textcolor}{}
{%
 \definecolor{BLACK}{gray}{0}
 \definecolor{WHITE}{gray}{1}
 \definecolor{RED}{rgb}{1,0,0}
 \definecolor{GREEN}{rgb}{0,1,0}
 \definecolor{BLUE}{rgb}{0,0,1}
 \definecolor{CYAN}{cmyk}{1,0,0,0}
 \definecolor{MAGENTA}{cmyk}{0,1,0,0}
 \definecolor{YELLOW}{cmyk}{0,0,1,0}
 }
\makeatother

\begin{document}

\title{The Power of One Qubit in Machine Learning}

\author{Roohollah Ghobadi}
\thanks{Corresponding author: farid.ghobadi@1qbit; \\
jaspreet.oberoi@1qbit.com, 
ehsan.zahedinejad@1qbit.com}
\affiliation{1QB Information Technologies (1QBit), Vancouver, BC, Canada}\affiliation{Departments of Computer Science and Mathematics, University of British Columbia, Vancouver, BC, Canada}
\author{Jaspreet S. Oberoi}
\affiliation{1QB Information Technologies (1QBit), Vancouver, BC, Canada}\affiliation{School of Engineering Science, Simon Fraser University, Burnaby, BC, Canada}
\author{Ehsan Zahedinejhad}
\affiliation{1QB Information Technologies (1QBit), Vancouver, BC, Canada}

\date{\today}

\begin{abstract}
Kernel methods are used extensively in classical machine learning, especially in pattern recognition. Here we propose a kernel-based quantum machine learning algorithm which can be implemented on a near-term, intermediate-scale quantum device. Our  method estimates classically intractable kernel functions, using a restricted quantum model known as ``deterministic quantum computing with one qubit''. Our work provides a framework for studying the role of quantum correlations other than quantum entanglement, for machine learning applications.  

\end{abstract}
\maketitle{}

{\it Introduction.} Noisy, intermediate-scale quantum (NISQ) devices, consisting of up to a few hundred qubits, have been presented as a new frontier for achieving ``quantum supremacy'', that is, surpassing the performance of classical computing devices~\cite{preskill}. This is plausible due to quantum advantages offered by non-universal quantum computational models such as Boson sampling~\cite{aaronson}, instantaneous quantum polynomial-time (IQP) sampling~\cite{bremner2016average}, and deterministic quantum computing with one qubit (DQC1)~\cite{Knill}. Recent experiments seeking quantum supremacy have involved  sampling from the output distribution of such non-universal models.

The potential applications of NISQ devices are a subject of extensive investigation in various fields, including quantum chemistry~\cite{lloyd1996universal,aspuru2005simulated,lanyon2010towards,KMT+17,matsuura2018vanqver}, quantum optimization~\cite{FGG+14}, and machine learning. Machine learning could benefit from NISQ devices due to the favourable exponential scaling of the Hilbert space and the potential capability of quantum correlations present in these devices to unveil hidden correlations in big data~\cite{Biamonte,perdomo2018opportunities}. 

Proposals for using NISQ devices for machine learning include quantum Boltzmann machines~\cite{AAR+18}, quantum clustering algorithms~\cite{OMA+17}, and quantum neural networks~\cite{FN18}. Very recently, a kernel-based supervised machine learning method---one based on a similarity measure between data points---has been proposed as an alternative route toward achieving a quantum advantage~\cite{schuld2019quantum,havlivcek2019supervised}. In the mentioned approach, a quantum processing unit is used to estimate a computationally expensive kernel function which can then be used as an input to a classical machine learning algorithm.

The DQC1 model is a non-universal quantum computing model which provides an exponential speedup in estimating the normalized trace of a unitary matrix, independent of the size of the matrix, over classical computing methods~\cite{Knill}. The quantum speedup achieved by the  DQC1 model is attributed to a non-classical correlation known as quantum discord ~\cite{Datta2,Datta}, which is robust against noise. It has also been shown that the DQC1 model cannot be efficiently simulated using classical devices unless the polynomial-time hierarchy collapses to the second level~\cite{fujii2018impossibility}. 

In this paper, we propose a QML algorithm that can be implemented on a NISQ device. Our scheme utilizes and redirects the computational advantage offered by the DQC1 model for estimating the kernels that are classically intractable to compute. We also provide a necessary condition for a kernel to be classically tractable. We then present simulation results for two synthesized datasets to demonstrate the efficacy of our method.

\vspace{1em}

{\it The kernel method.} To set the stage for our proposal, we introduce support vector machines (SVM) and the kernel method and  in the context of supervised machine learning. Let us assume a set of training ($X_\text{train}$) and test ($X_\text{test}$) datasets, where \mbox{$X=(X_\text{train}\cup{X_\text{text}})\subset{\mathbb{R}^d}$}. Each data point $\vec{x}\in{X}$ is assigned a label through a mapping $s:X\rightarrow{\{+1,-1\}}$. Classifying the data involves using training (i.e., labelled) data $X_\text{train}\rightarrow\{+1,-1 \}$ to find a classifier $f$ which can, with high probability, predict the correct label of the unseen (i.e., test) data points $X_\text{test}$ (i.e.,~$f:X_\text{test}\rightarrow\{+1,-1\}$).

\begin{figure}[t]
\includegraphics[width=0.46\textwidth]{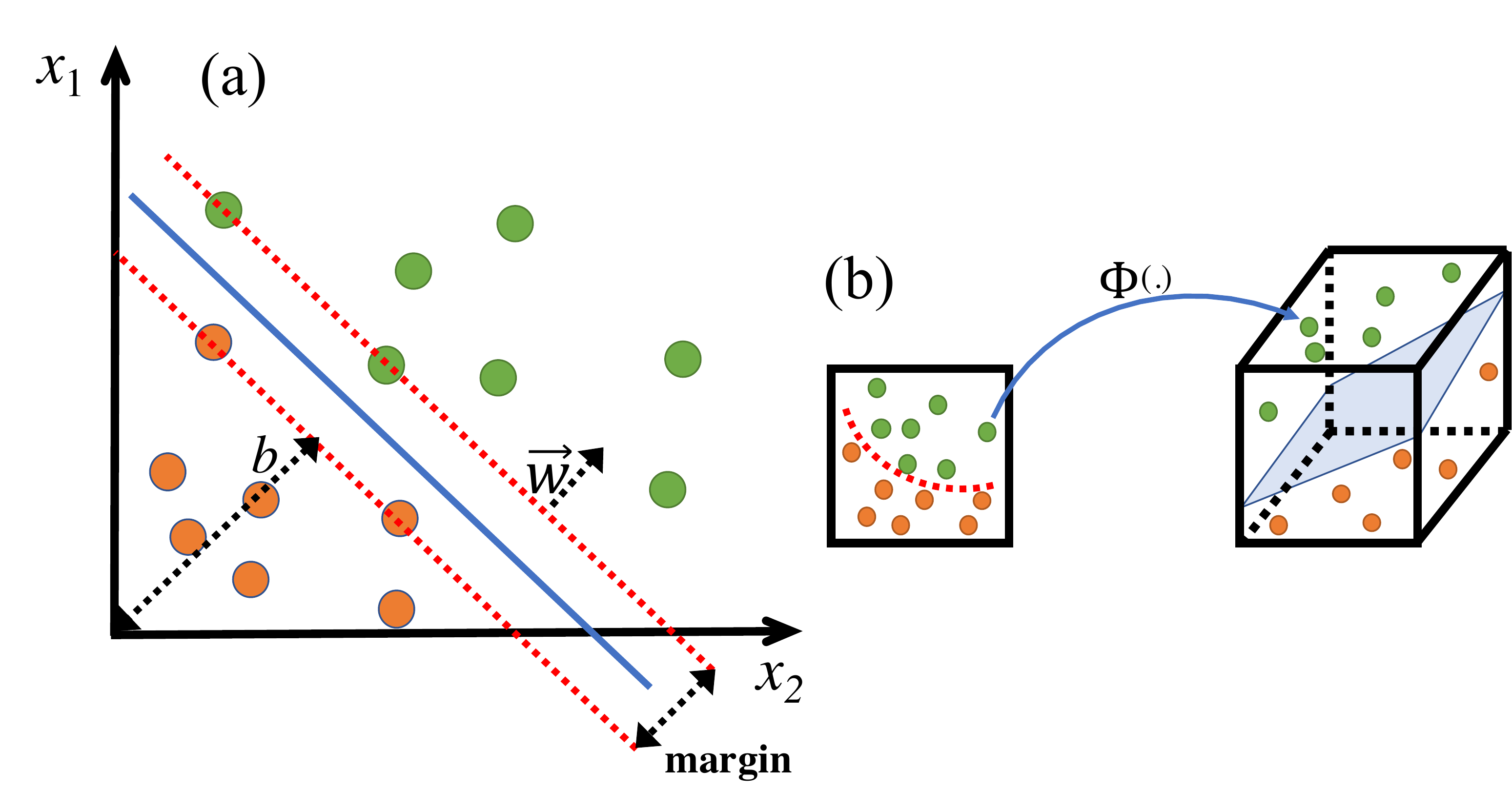}
\caption{(a) A support vector machine (SVM) with a two-dimensional linearly separable dataset. Circles in green and orange represent samples of two classes in the dataset. The support vectors (on the dotted red line) are those samples from each class that are closest to the decision boundary, represented by the blue line, with maximized margin. The variable $b$ is the offset and $\vec{w}$ is a normal vector to the decision boundary. (b) The kernel method: two non-linearly separable classes (left) can become linearly separable when mapped to a higher-dimensional space (right). 
} 
\label{kernel}
\end{figure}

For the simple case of a linearly separable dataset, as shown in Fig.~\ref{kernel}(a), one can find a hyperplane \mbox{$f(x)= \text{sign}(\vec{w}\cdot\vec{x}+b)$}, where $\vec{w}$ and $b$ are the hyperplane normal vector and offset, respectively, both of which can be determined using the training data.

The classification problem can be reduced to maximizing the margin (which is proportional to $||\vec{w}||^{-2}$) between the hyperplane and nearest data points, known as support vectors, subject to the condition that \mbox{$y_{i}(\vec{w}\cdot\vec{x}_{i}+b)\geq1$} (see Fig.~\ref{kernel}(a)). We can express the classifier in terms of the Lagrange multiplier as \mbox{$f(\vec{x})=\text{sign}(\sum_i\alpha_{i}y_{i} {\vec{x}^T\cdot\vec{x}_{i}})$}, with $\alpha_{i}\in\mathbb{R}$~\cite{hofmann2008kernel}. The classifier function depends on the inner product of the data points, which is the basis of the kernel method and the generalization of SVMs to nonlinear classifiers. To generalize the SVMs,
one can define the feature map that transforms the original data points into vectors in a higher-dimensional space, as shown in Fig.~\ref{kernel}(b). Formally, we define the feature map $\Phi:X\rightarrow\mathcal{H}$, where $\mathcal{H}$ is a Hilbert space. The kernel function, a similarity measure between data points $\vec{x},\vec{x}'\in X$, can be defined as  $K(\vec{x},\vec{x}')=\braket{\Phi(\vec{x})|\Phi(\vec{x}')}$, where the bra--ket notation shows the inner product on the Hilbert space $\mathcal{H}$.

The link between the kernel function and machine learning has been established by the ``representer theorem''~\cite{hofmann2008kernel}, which guarantees that for a positive semi-definite kernel 
\footnote{The kernel is positive semi-definite if $\forall   c_{i},c_{j}\in\mathbb{C}$, and $\forall \vec{x}_{i},\vec{x}_{j}\in X$, we have $\sum_{i,j}c_{i}c^{*}_{j}K(\vec{x}_{i},\vec{x}_{j})\geq0$} the classifier can be expressed as
\begin{equation}
f(\vec{x})=\sum_{i}\alpha_{i}K(\vec{x},\vec{x}_{i}),
\label{representer}
\end{equation}
where ~$\vec{x}\in{X}_{\text{test}}$, and $\vec{x}_i\in{X}_{\text{train}}$ are support vectors.

The kernel method can be extended to the quantum domain~\cite{havlivcek2019supervised,schuld2019quantum}. To do so, one can define the feature map $|\Phi(\vec{x})\rangle=U_{\Phi(\vec{x})}|0\rangle^{\otimes{n}}$, where  $\vec{x}$ is encoded in the quantum circuit $U_{\Phi(\vec{x})}$, and $|0\rangle^{\otimes{n}}$ is the initial state of $n$ qubits. The kernel function is then defined according to $K(\vec{x},\vec{x}')=\braket{\Phi(\vec{x})|\Phi(\vec{x}')}$. So long as it is possible to efficiently estimate a kernel using classical means, one cannot expect to attain a quantum advantage~\cite{schuld2019quantum,havlivcek2019supervised}. In other words, a necessary condition for achieving a quantum advantage in the kernel method is to realize a kernel function(s), one that is highly inefficient or intractable for classical devices to estimate~\cite{havlivcek2019supervised,van2006quantum}.

\vspace{1em}

{\it DQC1.} The deterministic quantum computing with one qubit (DQC1) model is a non-universal quantum computing model~\cite{Knill} which provides an exponential speedup in estimating the normalized trace of a unitary matrix, independent of the size of the matrix, over classical computing resources~\cite{Datta, fujii2018impossibility}. The model defies the common notion that achieving a quantum advantage in computation requires pure quantum states and quantum entanglement as resources. The DQC1 circuit is depicted in Fig.~2(a), where the initial state $|0\rangle\langle0|\otimes\rho_{n}$ evolves under the unitary interaction
\begin{equation}
 U=|0\rangle\langle0|\otimes \mathbb{1}_{n}+|1\rangle\langle1|\otimes U_{n},
 \label{unitary}
 \end{equation}
where $\mathbb{1}_n$ is the $N\times{N}$ ($N=2^n$) identity matrix and $U_n$ is a unitary operator acting on the $n$-qubit register. The final state $\rho_f$ of the control qubit, as in~\cite{Knill}, becomes 
\begin{equation}
\rho_{f}=\frac{1}{2}\begin{pmatrix}
1&\text{Tr}(\rho_{n}U_{n})\\
\text{Tr}(\rho_{n}U_{n}^{\dagger})&1
\end{pmatrix},
\label{rho1}
\end{equation}
where \text{Tr} refers to the trace of the matrix. In the special case where $\rho_n=\frac{\mathbb{1}_n}{N}$,  the off-diagonal terms in Eq.~(\ref{rho1})
become $\frac{1}{N}\text{Tr}(U_{n})$, suggesting that one can estimate the trace of an arbitrarily large matrix $U_{n}$ by measuring the decoherence of the control qubit. By measuring the Pauli operators of the control qubit, one obtains \mbox{$\langle\sigma_{x}\rangle=\frac{1}{N}\text{Re}[\text{Tr}(U_{n})]$} and $\langle\sigma_{y}\rangle=\frac{1}{N}\text{Im}[\text{Tr}(U_{n})]$.  The number of measurements required to estimate the trace within a distance $\epsilon$ with accuracy $\delta$ is given by $O(\log(1/\delta)/\epsilon^{2})$,  independent of the number of register qubits \cite{Knill}. Note that one can obtain the same ($\epsilon$, $\delta$) scaling for a special case where the $U_{n}$ is  a real, positive semi-definite matrix using a classical randomized algorithm \cite{avron2011randomized}. Importantly, replacing the pure control qubit with $\frac{\mathbb{1}_{1}+\beta\sigma_{z}}{2}$ (where $\beta\in\mathbb{R}$, and $\sigma_z$ denotes the Pauli-Z operator) adds an overhead of $\beta^{-2}$, suggesting that the quantum advantage of the model is robust against imperfect preparations of the control qubit.

It is worth noting that the DQC1 model has been experimentally realized for optical~\cite{lanyon2008experimental}, nuclear magnetic resonance~\cite{passante2011experimental}, and superconducting~\cite{Wang} qubits. In addition, a cold-atom-based scheme using the Rydberg interaction with 100 qubits has been proposed recently for DQC1~\cite{Mansell}.

\begin{figure}
    \centering
    \includegraphics[width=0.48\textwidth]{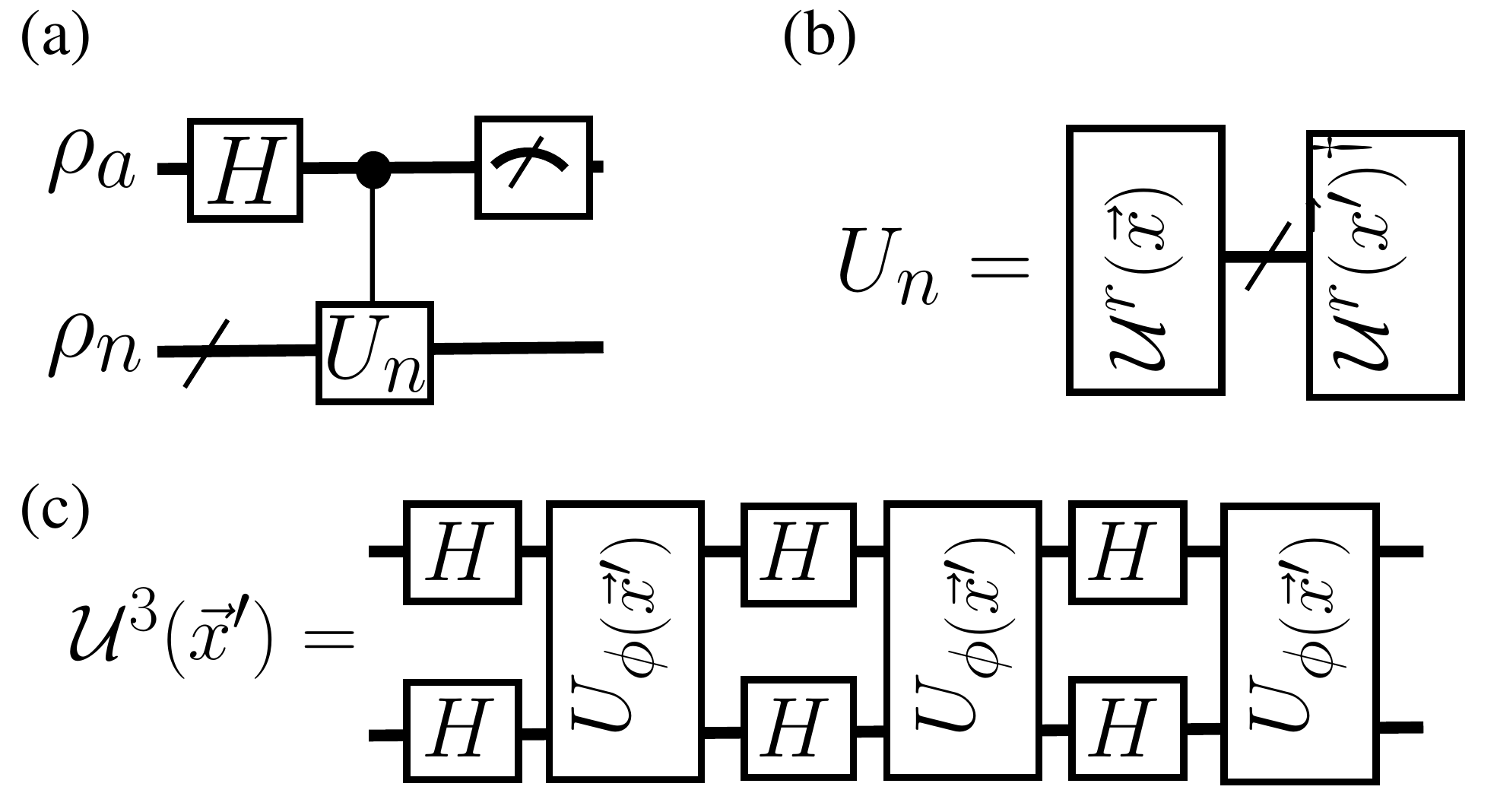}
    \caption{(a) The DQC1 circuit. The  control qubit is initialized in the  state $\rho{_a}=\ket{0}\bra{0}$, and the $n$-qubit register (indicated by a ``/'') is initialized in the state $\rho{_n}$. $H$ denotes the Hadamard gate and $U_n$ is a unitary operator acting on the register. (b)~Our implementation of a circuit with decomposition  \mbox{$U_{n}=\mathcal{U}^{r^\dagger}(\vec{x})\mathcal{U}^{r}(\vec{x}')$}, 
    where $\mathcal{U}^{r}$ is the encoding circuit with a depth of $r$. (c) The circuit structure of the unitary operator $\mathcal{U}$ adapted from~\cite{havlivcek2019supervised} to construct the kernel function for the two samples $\{\vec{x},\vec{x}'\}\in{X}$. We have chosen $r=3$ for our simulation. Here $\phi(\vec{x})$ is the feature encoding (defined in the text).}
    \label{DQC1}
\end{figure}

\vspace{1em}

{\it Method.} Here, we explain how we employ the DQC1 circuit to construct the kernel function. We consider a decomposition of $U_{n}$ in Eq.~(\ref{unitary}) as ~$U_{n}=\mathcal{U}^{r}(\vec{x})\mathcal{U}^{r^\dagger}(\vec{x}')$, where $\mathcal{U}^{r}(\vec{x})$ and $\mathcal{U}^{r^\dagger}(\vec{x}')$ represent the encoding of two data points $\vec{x}$ and $\vec{x}'$, respectively, and $r$ is the depth of the quantum circuit. We define the kernel function as

\begin{equation}
K(\vec{x},\vec{x}')=\text{Tr}(\rho_{n}\mathcal{U}^{r}(\vec{x})\mathcal{U}^{r^\dagger}(\vec{x}')).
\label{kernel2}
\end{equation}

Using~\eqref{rho1} and~\eqref{kernel2}, we obtain
\begin{equation}
\rho_{f}=\frac{1}{2}\begin{pmatrix}
1&K(\vec{x},\vec{x}')\\
K^{*}(\vec{x},\vec{x}')&1
\end{pmatrix},
\label{rho2}
\end{equation}
which is the main result of this work. Once the kernel has been obtained, one can use it in any kernel-based machine learning algorithm.

The flexibility in choosing $\rho_{n}$ and $\mathcal{U}^{r}$ in Eq.~(\ref{kernel2}) allows one to adapt this method to cater to different kernels, depending on the dataset. Our scheme can be applied to both discrete and continuous variable systems~\cite{lau2017quantum, liu2016power}. For example, using $\rho_{n}=|0\rangle\langle 0|^{\otimes n}$ and  $\mathcal{U}^{r}(\vec{x})=D(\vec{x})$ in Eq.~(\ref{kernel2}) with $D$ as the displacement operator~\cite{scully1999quantum}, one obtains the well-known, shift-invariant radial basis function (RBF) kernel $K(\vec{x},\vec{x}')=e^{-|\vec{x}-\vec{x}'|^{2}}$ (see also~\cite{chatterjee2016generalized}). As another example, for $\rho_{n}=\frac{\mathbb{1}_n}{N}$, the resulting kernel is \mbox{$K(\vec{x},\vec{x}')=\delta^{n}(|\vec{x}-\vec{x}'|^{2})$}, where $\delta$ is the Dirac delta function. 

Please note that shift-invariant kernels, such as the RBF kernel, can be efficiently estimated classically. To show this, one can use Bochner's theorem to write a shift-invariant kernel \mbox{$K(\vec{x}-\vec{x}')$}
as the Fourier transform of a probability distribution $p(\omega)$~\cite{rudin2017fourier}, that is, 

\begin{equation}
K(\vec{x}-\vec{x}')=\int d\omega p(\omega)e^{i\omega.(\vec{x}-\vec{x}')}.
\label{kshift}
\end{equation}

Since $|e^{i\omega.(\vec{x}-\vec{x})}|^{2} = 1$, Hoeffding's inequality guaranties an efficient estimation of Eq.~\eqref{kshift} with a maximum error of $\epsilon$ by drawing $O(\epsilon^{-2})$ samples from $p(\omega)$ (see also~\cite{rahimi2008random}). This argument applies to rotationally invariant kernels as well~\cite{rotation}. 

\vspace{1em}

\textit{Simulation.} We now provide a proof-of-principle example, in which a particular DQC1 quantum circuit performs the classification task on two  datasets. In~\cite{havlivcek2019supervised}, a quantum circuit is proposed, which has been conjectured to lead to a kernel that is intractable for a classical device (see also~\cite{van2006quantum}). We consider a circuit that has the same feature map as~\cite{havlivcek2019supervised},
\mbox{$\mathcal{U}^{r}(\vec{x})=\prod_{i=0}^{r}( \mathcal{U}_{\phi(\vec{x})}H^{\otimes n})_i$},
where $H$ denotes the Hadamard gate, and \mbox{$\mathcal{U}_{\phi(\vec{x})}=\text{exp}(i\sum_{S}\phi_{S}(\vec{x})\prod_{i}\sigma_{z}^{i})$}, with $\phi_{i}(\vec{x})=x_{i}$ for $i\in\{1,2\}$ and \mbox{$\phi_{1,2(\vec{x})}=(\pi-x_{1})(\pi-x_{2})$}. The requirement for obtaining a kernel that is non-translationally invariant imposes a lower bound on the depth of the circuit. For example, in the case of $r=1$, from Eq.~(\ref{kernel2}) we obtain $K(\vec{x}-\vec{x}')=\text{Tr}(\rho_{n}\mathcal{U}_{\phi(\vec{x}-\vec{x}')})$ and, therefore, the resulting kernal is classically simulatable (see also ~\cite{havlivcek2019supervised}). For this reason, in our simulation, we use $r=3$ to ensure that the kernel is not shift invariant.

\begin{figure}[t]
\includegraphics[width=0.45\textwidth]{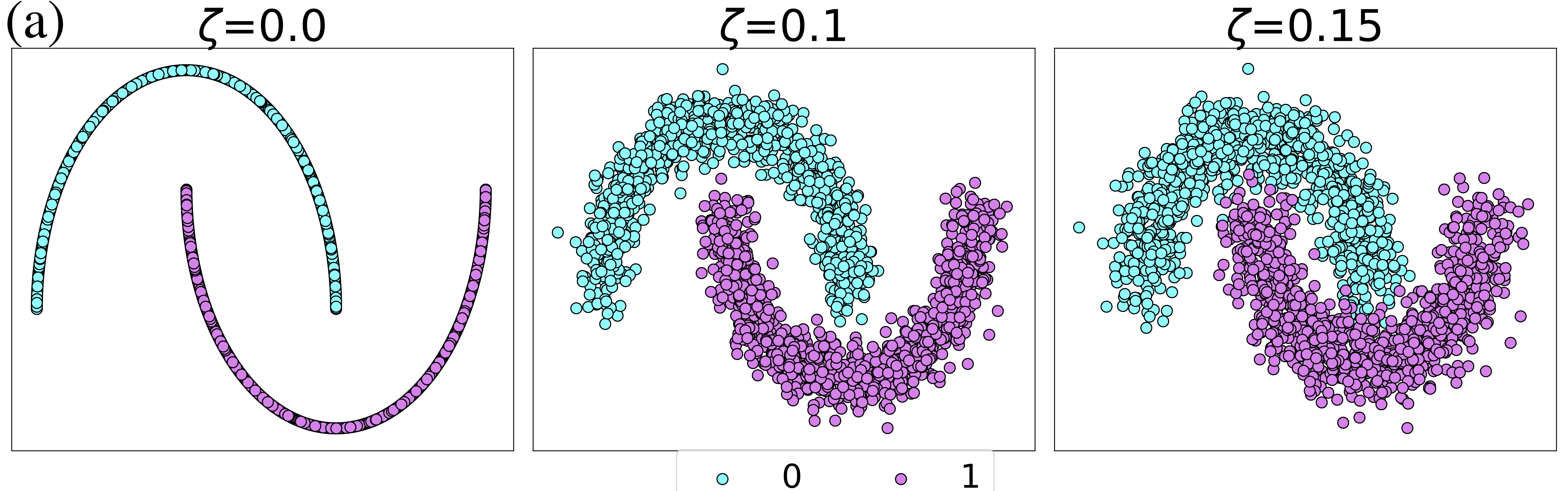}\\
\includegraphics[width=0.45\textwidth]{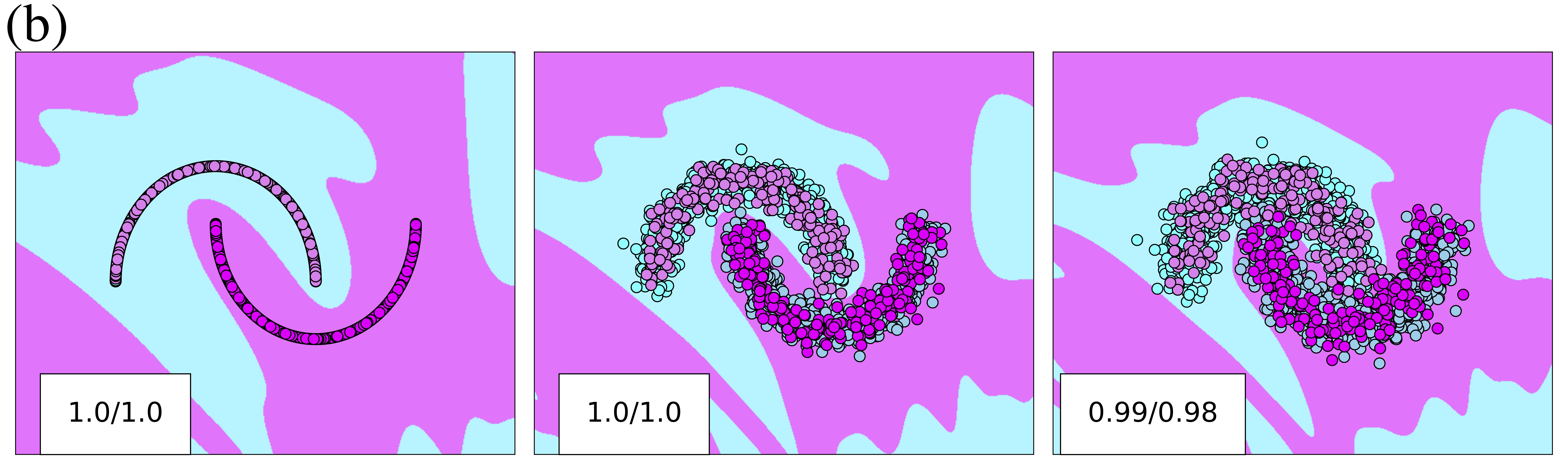}\\
\includegraphics[width=0.45\textwidth]{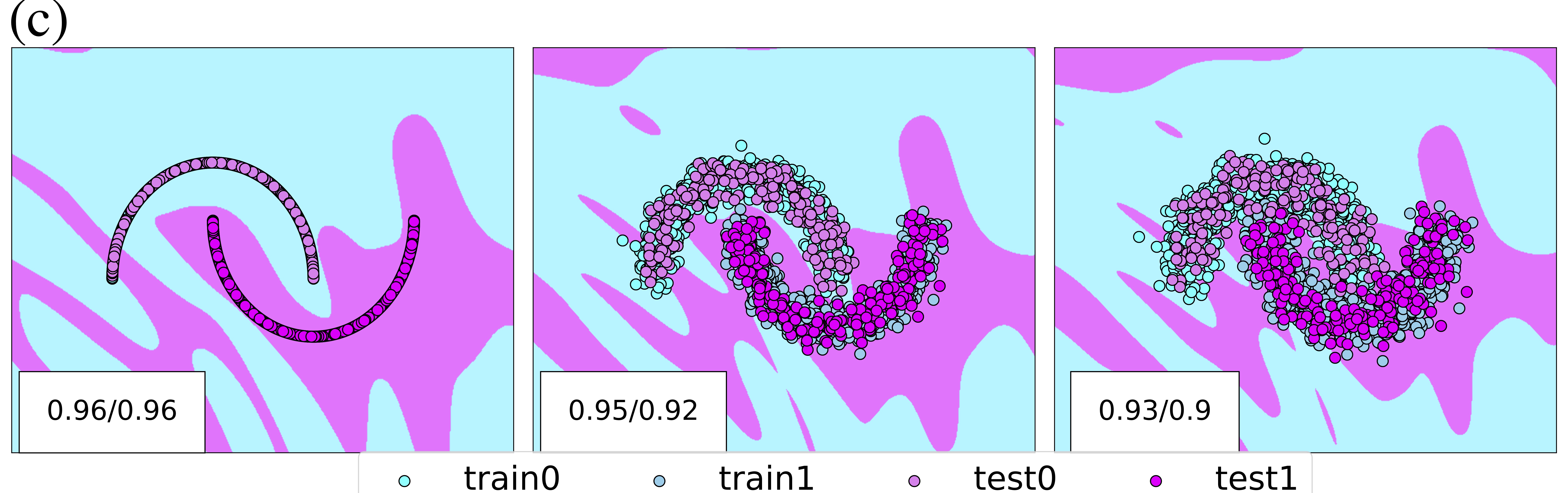}\\
\caption{(a) Plots of the synthesized moons datasets, with $2000$ samples, for a given noise (in the dataset) value \mbox{$\zeta = 0.0, 0.1, 0.15$}. We use the make\_moons method in the \mbox{scikit-learn} dataset module to generate each dataset. Shown are the training/test scores (i.e., accuracy) for (b)  registers in the mixed state and (c) registers in the pure state.}
\label{fig:moon_dataset}
\end{figure}

\begin{figure}[t]
\includegraphics[width=0.45\textwidth]{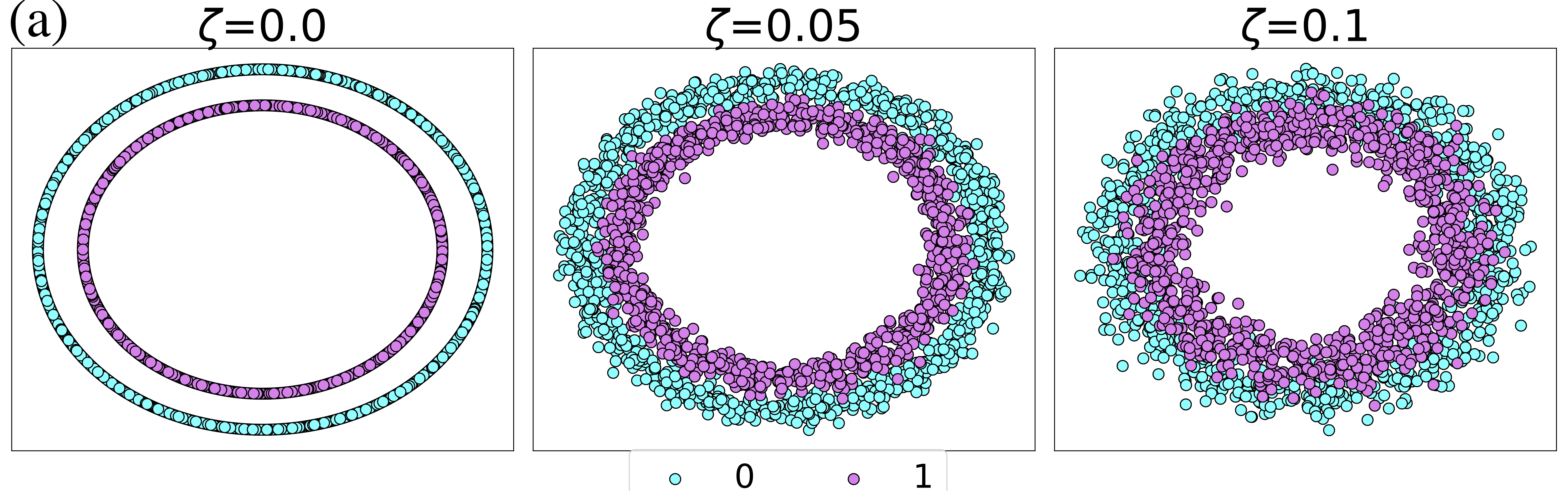}\\
\includegraphics[width=0.45\textwidth]{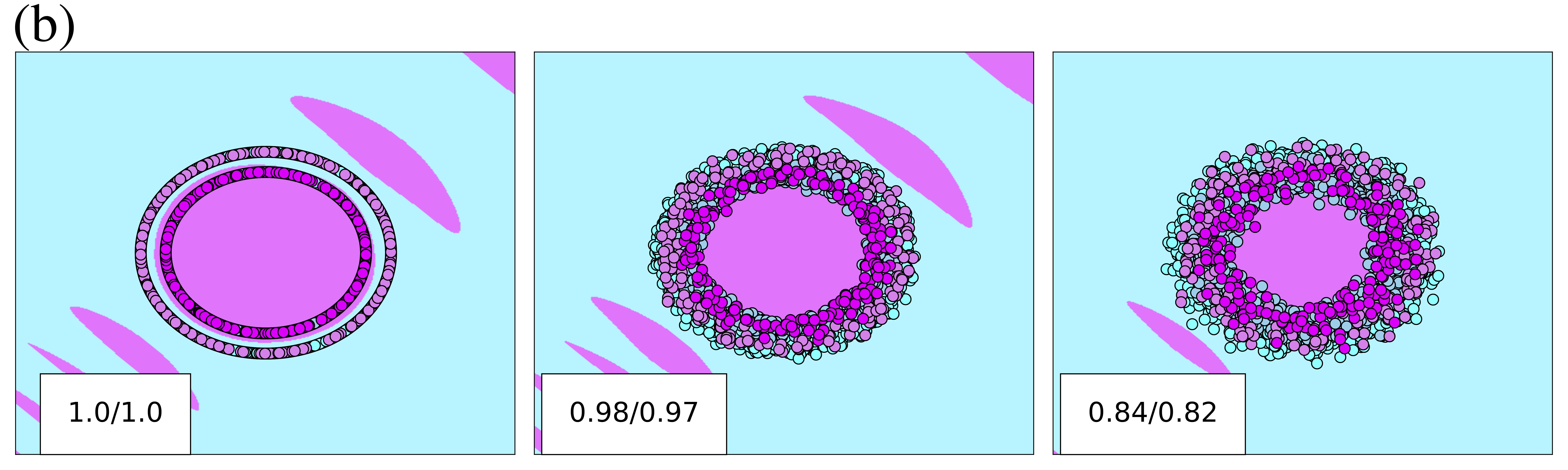}\\
\includegraphics[width=0.45\textwidth]{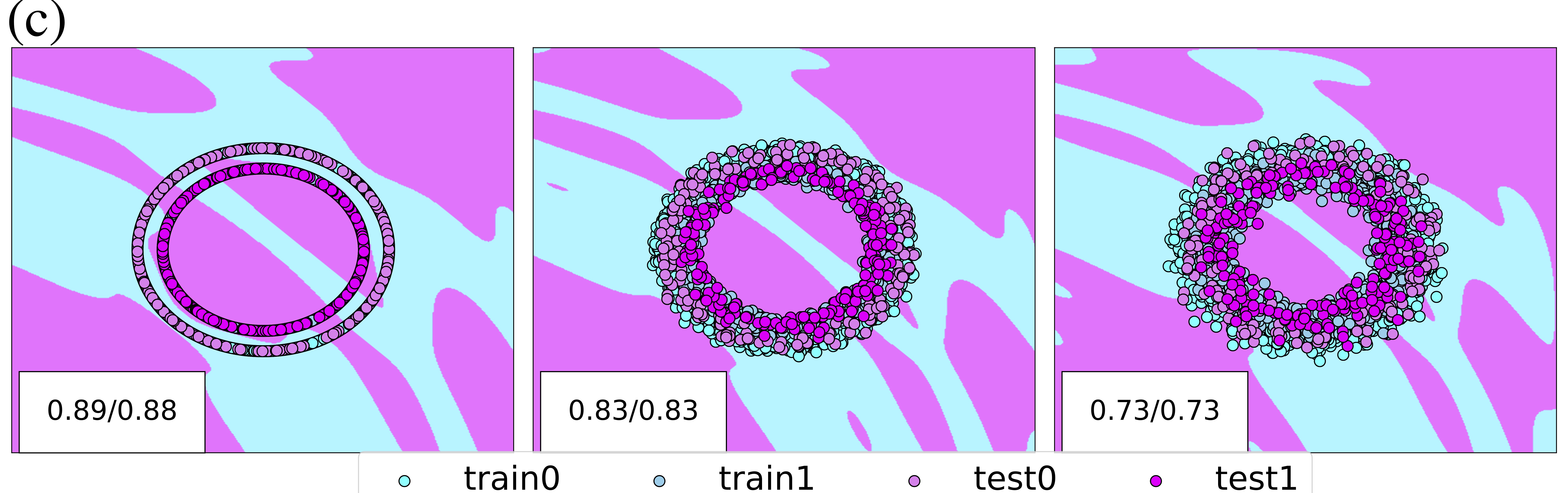}\\
\caption{{(a) Plots of the synthesized circles dataset, with $2000$ samples, for a given noise (in the dataset) value \mbox{$\zeta = 0.0, 0.05, 0.1$}. We use the make\_circles method in the \mbox{scikit-learn} dataset module to generate each dataset. Shown are the training/test scores (i.e., accuracy) for (b) registers in the mixed state and  (c) registers in the pure state.}}
\label{fig:circle_dataset}
\end{figure}

We run our experiments on two two-dimensional datasets with binary labels. The ``make\_moons'' and ``make\_circles'' methods in the ``scikit-learn'' datasets module are used to generate these datasets. We consider three levels of noise to generate three datasets for both the make\_moons and make\_circles methods. For the moons dataset, we use the noise values  $\zeta=0.0$, $0.1$, $0.15$ (see Fig.~\ref{fig:moon_dataset}(a)), and for the circles dataset, we use $\zeta=0.0$, $0.05$, $0.1$ (see Fig.~\ref{fig:circle_dataset}(a)). Consistently, across all six datasets, a total of $2000$ points are generated and each dataset is split into $1600$ training and $400$ test samples.

We run the quantum circuit in Fig.~\ref{DQC1}(a)--(c) for each pair of training data and estimate the kernel $K(\vec{x},\vec{x}')$ by directly calculating the trace of the $U_n$ operator in Fig.~\ref{DQC1}(a). We then use the absolute value of the resultant kernel to find the support vectors for the classifier. Finally, we use the classifier to predict the labels for the test data.

Figures~\ref{fig:moon_dataset}(b)--(c) and~\ref{fig:circle_dataset}(b)--(c) show the performance of our kernel approach on the ``moons'' and ``circles'' datasets. For each dataset, we report the classification accuracy on the ``training/test'' datasets. Figure~\ref{fig:moon_dataset}(a) shows the moons dataset, generated using three levels of noise. Figure~\ref{fig:moon_dataset}(b) shows the results of our quantum classifier for the fully mixed state $\rho_{n}=\frac{\mathbb{1}_n}{N}$, and Fig.~\ref{fig:moon_dataset}(c) shows the results when the initial state is the pure state $\rho_{n}=|0\rangle\langle0|^{\otimes n}$. It is interesting to note that the classification accuracy is reduced when we use the pure state initialization instead of the mixed state initialization. The performance of our algorithm is consistent across the two datasets. In both cases, we observe that the quantum circuit  learns to classify well (see Tables~\ref{table:moons} and~\ref{table:circles}), and also that the circuit prepared at the mixed state outperforms the one prepared at the pure state.

We also compare the performance of our quantum kernel approach with that of the classical RBF kernel. We consider the same training/test sizes as those used for the quantum case. Unlike the quantum kernel, for training the SVM using the classical RBF kernel, we have used five-fold cross-validation to ensure that we obtain the best results for the classical RBF kernel. Tables~\ref{table:moons} and~\ref{table:circles} provide a summary of the performance of the SVM using quantum and classical kernels, respectively. For both the moons and circles datasets, the classification results of the SVM using the quantum kernel with the mixed state are comparable to those of the SVM using the classical RBF kernel.

\begin{table}[]
\caption{A summary of the performance of the SVM algorithm using the quantum and classical kernels on the moons datasets (see Fig.~$\ref{fig:moon_dataset}$). We report the training/test score of the classifier when the register's qubits are initialized at mixed and pure states. The classical kernel is an RBF kernel. Here $\zeta$ denotes the noise values used in generating the datasets.}
\begin{tabular}{lccccccc}
\cline{2-7}
            & \multicolumn{2}{c}{$\zeta=0$} & \multicolumn{2}{c}{$\zeta=0.1$} & \multicolumn{2}{c}{$\zeta=0.15$} \\ \cline{2-7} 
            & training          & test         & training           & test          & training           & test           \\ \hline
mixed state & 1.0            & 1.0          & 1.0             & 1.0           & 0.99            & 0.98           \\ \hline
pure state  & 0.96           & 0.96         & 0.95            & 0.92          & 0.93            & 0.90           \\ \hline
RBF kernel  & 1.0            & 1.0          & 1.0             & 1.0           & 1.0             & 0.98           \\ \hline
\end{tabular}
\label{table:moons}
\end{table}

\begin{table}[]
\caption{A summary of the performance of the SVM algorithm using the quantum and classical kernels on the circles datasets (see Fig.~$\ref{fig:circle_dataset}$). We report the training/test score of the classifier when the register's qubits are initialized at mixed and pure states. The classical kernel is an RBF kernel. Here $\zeta$ denotes the noise values used in generating the datasets.}

\begin{tabular}{lccccccc}
\cline{2-7}
            & \multicolumn{2}{c}{$\zeta=0$} & \multicolumn{2}{c}{$\zeta=0.05$} & \multicolumn{2}{c}{$\zeta=0.1$} \\ \cline{2-7} 
            & training          & test         & training           & test          & training           & test           \\ \hline
mixed state & 1.0            & 1.0          & 0.98            & 0.97          & 0.84            & 0.82           \\ \hline
pure state  & 0.89           & 0.88         & 0.83            & 0.83          & 0.73            & 0.73           \\ \hline
RBF kernel  & 1.0            & 1.0          & 0.98            & 0.97          & 0.85            & 0.81           \\ \hline
\end{tabular}
\label{table:circles}
\end{table}

\vspace{1em}

{\it The effect of noise.} As a final remark, we wish to comment on how the effect of noise inherent to quantum gates may be characterized in our scheme. Note that in the absence of noise, we have $K(\vec{x},\vec{x})=1$. In practice, however, to take the noise into account, one must modify Eq.~(\ref{kernel2}) into the equation $\widetilde{K}(\vec{x},\vec{x}')=\text{Tr}(\rho_{n}\mathcal{\widetilde{U}}(\vec{x})\mathcal{\widetilde{U}}^{\dagger}(\vec{x}'))$, where $\mathcal{\widetilde{U}}$ denotes the noisy experimental implementation of $\mathcal{U}$. Note that \mbox{$\widetilde{K}(\vec{x},\vec{x})<1$}. Having access to $\widetilde{K}(\vec{x},\vec{x})$, by measuring the control qubit, one can efficiently estimate the average fidelity, a measure of the impact of noise, by using $F(\vec{x})=\frac{|\widetilde{K}(\vec{x},\vec{x})|^{2}+N}{N^{2}+N}$~\cite{nielsen2002simple,Poulin}. 

\vspace{1em}

\textit{Conclusion.} We have proposed a kernel-based scheme for QML, based on the DQC1 model. We have numerically tested our method to classify data points of two-dimensional synthesized datasets using a two-qubit circuit. Our work highlights the role of quantum correlations such as quantum discord in machine learning and benefits from the relationship between the fidelity of the process and the kernel function, to assess  the effect of  noise. Our method provides a framework for exploring the possibility of achieving quantum supremacy in machine learning, as it provides a means to efficiently estimate classically intractable kernels using NISQ devices.

\section{Acknowledgement} 
We thank Mark Schmidt for  insightful discussions and for reviewing a draft of the paper. R. Ghobadi appreciates the multiple discussions he had with with Artur Scherer. We greatly thank Marko Bucyk for reviewing and editing the manuscript. Partial funding for this work was provided by a Mitacs Elevate fellowship.


\begin{thebibliography}{}

\bibitem{preskill} J. Preskill, Quantum \textbf{2} 330 (2018).

\bibitem{aaronson} S. Aaronson and A. Arkhipov, \textit{The computational complexity of linear optics}, Proceedings of the 43rd annual ACM symposium on Theory of computing, 2011, San Jose (ACM, New York, 2011), p. 333.

\bibitem{bremner2016average} Michael J. Bremner, Ashley Montanaro, Dan J. Shepherd, Phys. Rev. Lett. \textbf{117}, 080501 (2016). 

\bibitem{Knill} E. Knill, R. Laflamme, Phys.Rev.Lett.\textbf{81} 5672 (1998).

\bibitem{lloyd1996universal} S. Lloyd, Science, \textbf{273} 1073-1078 (1996). 

\bibitem{aspuru2005simulated} A. Aspuru-Guzik, A. Dutoi, P. Love and M. Head-Gordon, Science, \textbf{309}, 1704(2005).

\bibitem{lanyon2010towards}Lanyon, Benjamin P and Whitfield, James D and Gillett, Geoff G and Goggin, Michael E and Almeida, Marcelo P and Kassal, Ivan and Biamonte, Jacob D and Mohseni, Masoud and Powell, Ben J and Barbieri {\it et al}.
Nat. chem \textbf{2}, 106 (2010).

\bibitem{KMT+17}  A. Kandala, A. Mezzacapo, K. Temme, M. Takita, M. Brink, J. M. Chow, J. M. Gambetta, Nature \textbf{549}, 242 (2017).

\bibitem{matsuura2018vanqver} S. Matsuura, T. Yamazaki, V. Senicourt and A, Zaribafiyan, arXiv:1810.11511 (2018).

\bibitem{FGG+14} E. Farhi, J. Goldstone, S. Gutmann, arXiv:1411.4028  (2014).

\bibitem{Biamonte}   J. Biamonte, {\it et al}. Nature \textbf{549}, 195 (2017).

\bibitem{perdomo2018opportunities} A. Perdomo-Ortiz, M. Benedetti, J. Realpe-Gómez, R. Biswas,  Quantum Science and Technology \textbf{3}, 030502 (2018).

\bibitem{AAR+18} M. H. Amin, E. Andriyash, J. Rolfe, B. Kulchytskyy, R. Melko, Phys. Rev. X 8, 021050 (2018).

\bibitem{OMA+17} J. Otterbach,  R. Manenti, N. Alidoust, A. Bestwick, M. Block, B. Bloom, S. Caldwell, N. Didier, E. S. Fried, S. Hong, {\it et.al.} arXiv:1712.05771 (2017). 

\bibitem{FN18} E. Farhi, H. Neven, arXiv:1802.06002  (2018).

\bibitem{schuld2019quantum} M. Schuld and N. Killoran, Phys. Rev. Lett. \textbf{122}, 040504  (2019).

\bibitem{havlivcek2019supervised} V. Havl{\'\i}{\v{c}}ek,A.D. C{\'o}rcoles, Antonio, K. Temme, A.W. Harrow, A. Kandala,  J. M. Chow and J. M. Gambetta Nature \textbf{567}, 209 (2019). 

\bibitem{Datta2} A. Datta, A. Shaji, C. M. Caves, Phys. Rev. Lett. \textbf{100}, 050502 (2008).

\bibitem{Datta} A. Datta, G. Vidal, Phys. Rev. A \textbf{75}, 042310 (2007).

\bibitem{fujii2018impossibility} K. Fujii, H. Kobayashi, T. Morimae, H. Nishimura, S. Tamate, and S. Tani, Phys. Rev. Lett. 120, 200502 (2018). 

\bibitem{hofmann2008kernel} T. Hofmann, B. Sch{\"o}lkopf, and A.J. Smola, The annals of statistics 1171 (2008).

\bibitem{van2006quantum} Van Dam, W., Hallgren, S. and Ip, L. Quantum algorithms for some hidden shift problems. SIAM J. on Computing \textbf{36}, 763 (2006).

\bibitem{avron2011randomized} H.Avron and S. Toledo, Journal of ACM (JACM) \textbf{58}, 8 (2011). 

\bibitem{lanyon2008experimental} B. P. Lanyon, M. Barbieri, M. P. Almeida, and A. G. White, Phys. Rev. Lett. \textbf{101}, 200501 (2008).

\bibitem{passante2011experimental} G. Passante, O. Moussa, D. A. Trottier, and R. Laflamme, Phys. Rev. A \textbf{84}, 044302 (2011).

\bibitem{Wang} W. Wang, B. Yadin, Y. Ma, J. Ma, Y. Xu, L. Hu, H. Wang, Y. P. Song, Mile Gu, L. Sun, arXiv:1806.05543 (2018).

\bibitem{Mansell} C. Mansell, S. Bergamini, New Journal of Physics, \textbf{16}, 053045 (2014).

\bibitem{lau2017quantum}  H. K. Lau, R. Pooser, G. Siopsis, and C. Weedbrook, Phys. Rev. Lett. \textbf{118}, 080501 (2017).

\bibitem{liu2016power} N. Liu, J. Thompson, C. Weedbrook, S. Lloyd, V. Vedral, M. Gu, K. Modi, Phys. Rev. A \textbf{93}, 052304 (2016).

\bibitem{chatterjee2016generalized} R. Chatterjee and T. Yu, Quantum Information and Communication \textbf{17}, 1292 (2017).

\bibitem{scully1999quantum} M. O. Scully and M. S. Zubairy, Quantum optics (1999).

\bibitem{rudin2017fourier} W. Rudin, {\it Fourier analysis on groups}  (Courier Dover Publications 2017).

\bibitem{rahimi2008random} A. Rahimi, and R. Benjamin, {\it Advances in neural information processing systems} (2008) pp.1177-1184.

\bibitem{rotation} B. Bullins, C. Zhang, Yi Zhang, arXiv:1710.10230 , (2018).

\bibitem{nielsen2002simple} M. A. Nielsen, Phys. Lett. A \textbf{303} (4): 249 (2002). 

\bibitem{Poulin} D. Poulin, R. Blume-Kohout, R. Laflamme, H. Ollivier, Phys. Rev. Lett. \textbf{92}, 177906 (2004).

\end{thebibliography}
\end{document}